\documentclass[aps,prl,nofootinbib,amssymb,amsmath,graphix,twocolumn]{revtex4-1}
\pdfoutput=1
\usepackage{verbatim,graphics,graphicx,color,slashed}
\usepackage{ulem} 
\usepackage{hyperref}

\begin{document}

\title{$\nu \Lambda$MDM: A Model for Sterile Neutrino and Dark Matter Reconciles \\ 
Cosmological and Neutrino Oscillation Data after BICEP2}
\author{P. Ko}
\author{Yong Tang}
\affiliation{School of Physics, Korea Institute for Advanced Study,
 Seoul 130-722, Korea }

\begin{abstract}
We  propose a {\textit{ultraviolet}} complete theory for cold dark matter(CDM) 
and sterile neutrinos that can accommodate both cosmological data and neutrino oscillation experiments within $1\sigma$ level. We assume a new $U(1)_X$ dark gauge symmetry 
which is broken at $\sim \mathcal{O}$(MeV) scale resulting light dark photon.  
Such a light mediator for  DM's self-scattering and scattering-off sterile neutrinos  
can resolve three controversies for cold DM on small cosmological scales:  
{\textit {cusp vs. core}}, {\textit{too-big-to-fail}} and {\textit{missing satellites}}. 
We can also accommodate  $\sim O(1)$ eV scale sterile neutrinos as the hot dark 
matter(HDM) 
and can fit some neutrino anomalies from neutrino oscillation experiments within $1\sigma$. Finally the right amount of HDM can make a sizable contribution to dark radiation,  
and also helps to reconcile the tension between the data on the tensor-to-scalar 
ratio reported by Planck and BICEP2 Collaborations. 
\end{abstract}

\maketitle

\section{Introduction}
The standard model of cosmology, the so-called $\Lambda$CDM with the minimal 
six parameters, can explain well a wide range of cosmological observations, 
such as primordial abundance of light elements, cosmic microwave background(CMB) 
and large scale structures(LSS). Meanwhile, there are still some hints that indicate 
that new physics beyond the minimal $\Lambda$CDM model maybe is needed in order 
to explain  CDM sector better.

There are three controversies for CDM paradigm on small cosmological scales, 
{\textit {cusp vs. core}}, {\textit{too-big-to-fail}} and {\textit{missing satellites}}
(see Ref.~\cite{Weinberg:2013aya} for a review), which have triggered both astrophysical  
\cite{Efstathiou:1992sy, Pfrommer:2011bg, Silk:2010aw, Shapiro:2003gxa, MacLow:1998wv, Hopkins:2011rj, Pontzen:2011ty, VeraCiro:2012na, Wang:2012sv} and DM-related  investigations~\cite{Spergel:1999mh, Bode:2000gq, Dalcanton:2000hn, Zentner:2003yd,Sigurdson:2003vy,Kaplinghat:2005sy, Borzumati:2008zz, Boehm:2000gq, Kaplinghat:2000vt, Hu:2000ke,Lovell:2011rd, Feng:2009hw, Buckley:2009in, Loeb:2010gj, Vogelsberger:2012ku, Tulin:2013teo, Ko:2014nha}. 
A solution that resolves simultaneously these controversies  has been proposed in Ref.~\cite{Aarssen:2012fx}, where both DM and active neutrino interact with a new gauge boson 
with mass round $\mathcal{O}(\textrm{MeV})$. Then the  DM's velocity-dependent 
self-scattering cross section can be around $1$cm$^2/$g at the Dwarf satellites scale, and 
evades the constraints from Milky Way galaxy and galaxy cluster. Thus one can resolve 
the first two controversies. Meanwhile, the DM's scattering off cosmic neutrino background 
leads to its late kinetic decoupling at temperature 
$T_{\textrm{kd}} < \mathcal{O}(\textrm{keV})$, which is translated into a cut-off of 
the smallest protohalo mass $M_\textrm{cut}\sim \mathcal{O}(10^9)M_{\odot}$, 
resolving  the 3rd puzzle, namely missing satellites problem. However, since active neutrino couples to a MeV particle, such scenario is restrictively constrained~\cite{Ahlgren:2013wba,Shoemaker:2013tda,Wilkinson:2014ksa}.

The CMB data indicates that a small amount of relativistic species or 
hot dark matter(HDM) could exist at CMB time~\cite{Burenin:2013wg, Wyman:2013lza, Hamann:2013iba, Battye:2013xqa, Gariazzo:2013gua}, in addition to the standard 
three generations of active neutrinos. 
This is often parametrized as the effective number of additional neutrino species 
$\Delta N^{\textrm{cmb}}_\textrm{eff}$. 
It has been shown in Ref.~\cite{Hamann:2013iba} that the best fit 
 to all available data is given by 
\begin{equation}
\Delta N^{\textrm{cmb}}_\textrm{eff}=0.61\pm 0.30,\;
m^{\textrm{eff}}_{\textrm{hdm}}=(0.47\pm 0.13)\textrm{ eV}.
\end{equation}   
where $m^{\textrm{eff}}_{\textrm{hdm}}$ is the effective HDM mass.
Also,  it was shown very recently that a similar amount of HDM can help to relieve 
the tension of tensor-to-scalar ratio ($ \equiv r$) between Planck data~\cite{planck} 
and the recently announced measurement of B-mode polarization by BICEP2~\cite{bicep2}, without a running spectral index~\cite{Giusarma:2014zza,Zhangxin,Hu}. 

It is well known that sterile neutrino can serve as a HDM component of the 
universe. Sterile neutrino is also well motivated in order to solve accelerator~\cite{Athanassopoulos:1996jb,Aguilar:2001ty}, reactor~\cite{Mention:2011rk} 
and gallium anomalies~\cite{Abdurashitov:2005tb,Giunti:2012tn} in neutrino oscillation experiments.  Both reactor and gallium anomalies prefer a new mass-squared difference, 
$\Delta m^2 \gtrsim 1~\text{eV}^2$~\cite{Abazajian:2012ys}, while  accelerator experiments~\cite{Armbruster:2002mp,Aguilar-Arevalo:2013pmq,Antonello:2012pq} 
prefer $\Delta m^2 \sim 0.5~\text{eV}^2$. In all three cases the favored mixing angles 
are around $\sin^2{2\theta}\sim 0.1$.  Such a large mixing angle would in general 
lead to fully thermalized sterile neutrinos by oscillation and thus an increase 
of $\Delta N^{\textrm{cmb}}_\textrm{eff}=1$ for each sterile neutrino. This is in some tension with the above cosmological data Eq.~(1), as shown in global fit~\cite{Hamann:2011ge} including BBN, CMB and LSS data. 

The above tension can be relieved by introducing new interaction for sterile 
neutrino. The new interaction can generate a matter potential $V_{\textrm{eff}}$ 
that results in a tiny effective mixing angle $\theta_{m}$ in  matter~\cite{Akhmedov:1999uz} 
for $V_{\textrm{eff}}\gg \frac{\Delta m^2}{2E}$, 
\[
\sin^2{\theta_m}=\frac{\sin^2{2\theta_0}}{\left( \cos 2\theta_0 +\frac{2E}{\Delta m^2}V_\textrm{eff}\right)^2 + \sin^2 2\theta_0},
\]
where $\theta_0$ is the mixing angle in vacuum. 
As a result, the thermalization of sterile neutrino by oscillation can be suppressed 
and $\Delta N_\textrm{eff} < 1$ is easily obtained~\cite{Hannestad:2013ana,Dasgupta:2013zpn}. 
Recently it has been shown that the tension in the data can be reconciled at $2\sigma$ 
level within an {\textit{effective}} theory~\cite{Bringmann:2013vra} where a 
dim-5 operator is responsible for the active-sterile neutrino mixing.

In this paper, we propose a {\textit{ultraviolet}} complete theory for DM and sterile neutrino that can accommodate the aforementioned cosmological data and neutrino oscillation experiments within $1\sigma$ level. The model includes both CDM and HDM, and we call it  
the $\nu \Lambda$MDM(the first M stands for mixed). 
%

\section{Model for CDM and Sterile Neutrino}

We consider the standard seesaw model with two right-handed (RH) neutrinos 
(gauge singlet) $N_i(i=1,2)$\footnote{We could add more heavy $N$ in the Lagrangian~\ref{eq:lag} for leptogenesis~\cite{leptogenesis}, which will not affect our discussions in the following.},
and add a dark sector with $U(1)_X$ gauge symmetry and coupling $g_X$, and  
dark photon field $\hat{X}_\mu$,  and dark Higgs field $\phi_{X}$  and two different 
Dirac fermion $\psi$ and $\chi$ in the dark sector. 
All the new fields are SM gauge singlets. We assign equal $U(1)_{X}$ charges 
to $\phi_X$ and $\psi$, which is 
normalized to $1$. Then the most general gauge invariant renormalizable Lagrangian 
is given by 
\begin{widetext}
\begin{align}
\mathcal{L} 
   =& \mathcal{L}_{\textrm{SM}}+\bar{N}_{i}i\slashed{\partial}N_{i} - \left(\frac{1}{2}m_{ij}^{R}\bar{N}_{i}^{c}N_{j} + y_{\alpha i}\bar{L}_{\alpha}HN_{i}+h.c\right) - \frac{1}{4}\hat{X}_{\mu\nu}\hat{X}^{\mu\nu}
   -\frac{1}{2}\sin\epsilon\hat{X}_{\mu\nu}\hat{B}^{\mu\nu}
    \nonumber\\
    &{} +\bar{\chi}\left(i\slashed{D}-m_{\chi}\right)\chi+\bar{\psi}\left(i\slashed{D}-m_{\psi}\right)\psi + D_{\mu} \phi_{X}^{\dagger}D^{\mu}\phi_{X} - \left(
     f_{i}\phi_{X}^{\dagger}\bar{N}^c_{i}\psi + g_i \phi_{X}\bar{\psi}N_i+h.c\right)
    \nonumber\\
    &{} -\lambda_{\phi}\left[\phi_{X}^{\dagger}\phi_{X}-\frac{v_{\phi}^{2}}{2}\right]^{2} - \lambda_{\phi H}\left[\phi_{X}^{\dagger}\phi_{X} - \frac{v_{\phi}^{2}}{2}\right]\left[H^{\dagger}H-\frac{v_{h}^{2}}{2}\right],
\end{align}\label{eq:lag}
\end{widetext}
where $L_{\alpha}$ are the SM left-handed lepton doublets, $H$ is the SM 
Higgs doublet, and $\hat{B}$ is the field strength for SM $U(1)_{Y}$. The covariant 
derivative on a field $K$ is defined as
\[
D_\mu K = (\partial_\mu - iQ_K g_X \hat{X}_\mu)K \ \  ({\rm with} \ K=\chi,\psi,\phi_X ) \ .
\]
We have chosen the $U(1)_X$ charge for $\chi$ in such a way that the 
$\phi_X \bar{\chi} N_i$ term is forbidden by $U(1)_X$ gauge symmetry 
(otherwise $\chi$ may decay if kinamatically allowed). Thus $\chi$ would be stable 
and the DM candidate.
 
The local gauge symmetry is broken by the following vacuum configurations: 
\begin{eqnarray}\label{eq:vacuumstate}
\langle H\rangle=\frac{1}{\sqrt{2}}\left(\begin{array}{c}
0\\
v_{h}
\end{array}\right),\;\langle\phi_{X}\rangle=\frac{v_{\phi}}{\sqrt{2}},
\end{eqnarray}
where $v_h\simeq 246$GeV and $v_{\phi}\sim \mathcal{O}$(MeV) for our interest.
There will be  mixings among various  fields after the spontaneous gauge symmetry 
breaking.   The gauge kinetic mixing term results in tiny mixings among the physical gauge 
fields,  $A_\mu, Z_\mu$ and $X_\mu$. Also there is a mixing between Higgs fields $h$ 
and $\phi$  with 
 \[
 H\rightarrow\dfrac{v_h+h}{\sqrt{2}}~~ {\rm and}~~~ 
 \phi_X\rightarrow\dfrac{v_\phi+\phi}{\sqrt{2}}. 
\] 
Two scalar excitations  $h$ and $\phi$ can be expressed in terms of mass eigenstates, 
$H_1$ and $H_2$, as
\begin{eqnarray}
 h & = & H_1\cos \alpha  - H_2\sin \alpha,\\ 
 \phi & = & H_1 \sin \alpha + H_2\cos \alpha,
\end{eqnarray}
with a mixing angle $\alpha$. Because of the Higgs portal interaction ($\lambda_{\phi H}$ 
term) and the additional scalar $\phi$, the electroweak vacuum could be stable up to 
Planck scale without additional new physics beyond the particle contents presented in 
Eq.~(2) (see Refs.~\cite{Baek:2013qwa} for example).

A novel feature of this model is that there can be mixing among three active 
neutrinos $\nu_\alpha$, sterile neutrinos $N_i$ and dark fermion $\psi$ due to 
$y_{\alpha i}\bar{L}_{\alpha}HN_{i}$, $f_{i}\phi_{X}^{\dagger}\bar{N}_{i}\psi$ and 
$g_i \phi_{X}\bar{\psi}N_i$ after spontaneous gauge symmetry breaking. 
In order to correctly explain the active neutrino oscillation data, 
at least two $N$'s are needed, in which case two of $\nu_a$ are massive and the other one 
is massless. Then after diagonalization of $7\times 7$ mass matrix for $\nu_\alpha, N_i$ and $\psi$,  mass eigenstates are composed of 7 Majorana neutrinos, 
$\nu_a(a=1,2,3)$ and $\nu_{si}(i=4,...,7)$, or collectively $\nu_i=\nu_{iL} + \nu^c_{iR}$:
\begin{equation*}
\left(
\begin{array}{c}
\nu_\alpha \\
 N^c_i \\
\psi_L \\
\psi_L^c 
\end{array}
\right)
=U 
\left(
\begin{array}{c}
\nu_a \\
\nu_{s4} \\
\vdots \\
\nu_{s7} 
\end{array}
\right)_L, \;
U^{T}\mathbb{M}U=
\left(
\begin{array}{ccc}
m_1 & 		 & \\
 	& \ddots & \\
	&		 & m_7 	
\end{array}
\right),
\end{equation*}
where $U$ is the unitary mixing matrix that diagonalizes the mass matrix $\mathbb{M}$, 
\begin{equation*}
\mathbb{M}=
\left(
\begin{array}{ccc}
0_{3\times 3} & \dfrac{v}{\sqrt{2}}[y_{\alpha i}]_{3\times 2} & 0_{3\times 2} \\
\dfrac{v}{\sqrt{2}}[y_{\alpha i}]^T_{2\times 3} & \left[m^{R}_{ij}\right]_{2\times 2} &  \dfrac{v_\phi}{\sqrt{2}}(f_i \; g_i)_{2\times 2}\\
0_{2\times 3} &\dfrac{v_\phi}{\sqrt{2}}(f_i \; g_i)^T_{2\times 2} & 
\left(
\begin{array}{cc}
0 & m_\phi \\
m_\phi & 0
\end{array}
\right)
\\
\end{array}
\right).
\end{equation*}
In the following discussion, if not specified, we shall use $\nu_a$ and $\nu_s$ to collectively denote three active neutrinos and four sterile neutrinos, respectively.

The mixing also distributes the new $U(1)_X$ gauge/Yukawa interaction to all neutrinos with actual 
couplings depending on the exact mixing angles. We assume that the mixing angles between 
$\nu_\alpha$ and $\psi$ are negligible, compared to the mixing between $N_i$ and $\psi$. 
This can be easily achieved by adjusting $y_{\alpha i}$'s, $f_i$'s and $g_i$'s. 
A more straightforward way is to work in the flavor basis, in which only $N_i$ and 
$\phi$ have dark Yukawa and gauge interactions, respectively. Because of the new dark 
interactions for $\nu_s$, all sterile neutrinos $\nu_s$'s are not thermalized by oscillation 
from active neutrinos and thus can contribute to the number of effective neutrino 
by a proper amount, 
$\Delta N_{\textrm{eff}} < 1$ after BBN~\cite{Hannestad:2013ana,Dasgupta:2013zpn}.

The exact mass spectrum and mixing angles for $\nu_s$ are free, subject to conditions 
for fitting the data. We shall take at least one $\nu_s$ is around 1~eV and others 
as free, lighter or heavier, and the mixing angles among $\nu_s$ are large enough for  suppressing their production by oscillation from active neutrino.

Based on a different setup, our model improves a similar attempt presented in a recent 
paper~\cite{Bringmann:2013vra} in two aspects.
First, our model is renormalizable and thus ultraviolet complete, while the model in 
Ref.~\cite{Bringmann:2013vra} assumed a dim-5 operator for generating the active-sterile 
neutrino mixing and therefore depends on the UV completion. 
Second, we shall show below that the model presented in the present paper can reconcile 
the current cosmological data with neutrino oscillation experiments within $1\sigma$ 
rather than only within $2\sigma$ as discussed in \cite{Bringmann:2013vra}.
 
\section{Thermal History and CDM controversies}
\begin{figure}
\includegraphics[width=0.23\textwidth]{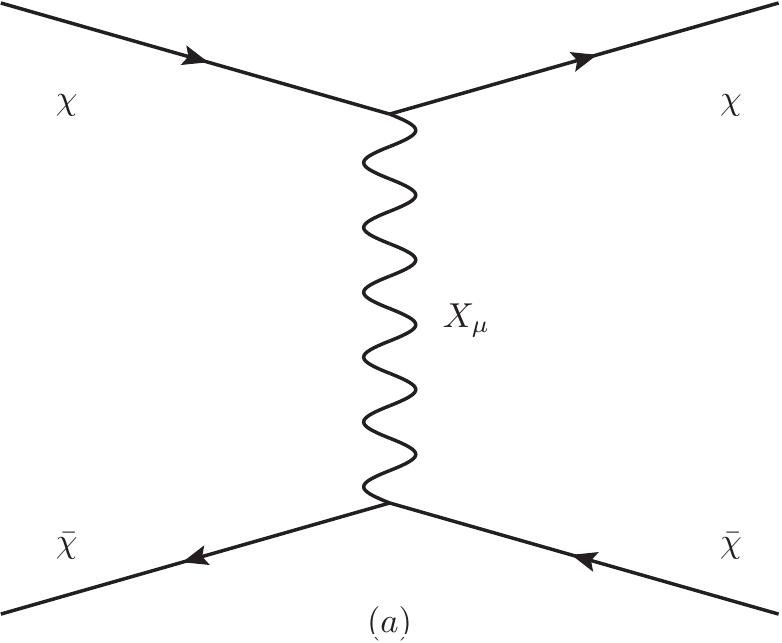}
\includegraphics[width=0.23\textwidth]{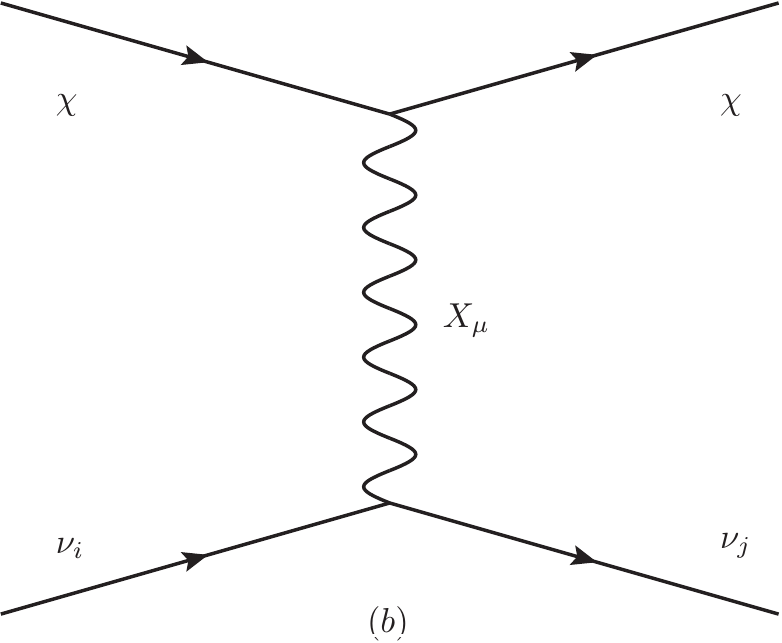}
\caption{Feynman diagrams for (a)$\chi \bar{\chi}$ and (b)$\chi \nu_s$ scattering where $i\neq j$ for $\nu_i$'s Majorana nature, $\bar{\nu}_i\gamma^\mu\nu_i=0$.}
\label{fig:DMnu}
\end{figure}

Communication between dark sector and SM particles or thermal history before BBN time 
is determined mostly by two mixing parameters, $\sin\epsilon$ and $\lambda_{\phi H}$.
$\sin\epsilon$ is contrained by DM direct searches around $ \sin\epsilon<10^{-9}$ for 
$\mathcal{O}$(TeV) $\chi$ and $\mathcal{O}$(MeV) $X_\mu$~\cite{Batell:2009vb}. 
And $\lambda_{\phi H}$ as small as $10^{-8}$ would be enough to thermalize the dark 
sector at $T\sim $TeV~\cite{decayvdm}.
After the cross sections of dark particles' scattering off SM particle drop below the expansion rate of the Universe, the dark sector decouples from the thermal bath of the visible sector 
and entropy density would be conserved separately in each sector.
The decoupling temperature of the dark sector, $T_x^{\textrm{dec}}$, 
would determine how much $\Delta N_{\textrm{eff}}$ is left at a later time. The exact value for 
$\Delta N_{\textrm{eff}}$ will be given in the following.  
 
Chemical decoupling of DM from the heat bath sets its relic density today.
After the temperature drops below $m_\chi$, $\chi$ starts to leave the chemical equilibrium 
and would finally freeze out at $T\simeq m_\chi/25$. To account for the correct thermal 
relic density,  the thermal cross section for $\chi \bar{\chi}$ annihilation 
$\langle \sigma v\rangle$ should be around $3\times 10^{-26}$cm$^3/$s. 
The dominant annihilation channel in this model is $\chi \bar{\chi} \rightarrow X_\mu X_\mu $,  
and the relic density requires the gauge coupling $g_X$ to be~\cite{Feng:2010zp}
\begin{equation}
g_X \sim \frac{0.50}{Q_\chi} \times \left(\frac{0.114}{\Omega_{\textrm{cdm}}}\right)^{\frac{1}{4}}\left(\frac{m_\chi}{\textrm{TeV}}\right)^{\frac{1}{2}},
\end{equation}
where $Q_\chi$ is the $U(1)_X$ charge of $\chi$ and shall be taken $\sim O(1)$ 
for definiteness in later discussion.
We shall focus on the CDM $\chi$ with mass $\sim$ TeV,  
which is preferred region as shown in Ref.~\cite{Aarssen:2012fx}.

Kinetic decoupling of $\chi$ from $\nu_s$ happens at much later time when the elastic 
scattering rate for $\chi \nu_s \leftrightarrow \chi \nu_s$ drops below some value determined 
by Hubble parameter $H$. The Feynman diagram is shown in Fig.~\ref{fig:DMnu}(b). For a thermal distribution of sterile neutrino, the decoupling 
temperature is given by 
\begin{equation}
T^{\textrm{kd}}_\chi \simeq 1 \textrm{keV}\left(\frac{0.1}{g_X}\right)
\left(\frac{ T_{\gamma} }{ T_{\nu_s} } \right)_{\textrm{kd}}^{\frac{3}{2}}
\left(\frac{ m_\chi }{ \textrm{TeV} } \right)^{\frac{1}{4}}
\left(\frac{ m_{X} }{ \textrm{MeV} } \right),
\end{equation}
where $T_\gamma$ and $T_{\nu_s}$ are the temperatures of CMB and sterile neutrinos, respectively. Except that DM is dominantly scattering off sterile neutrinos in our model rather than active ones, the above formula is similar to the one in Ref.~\cite{Aarssen:2012fx} and gives the approximate order-of-magnitude estimation, although the precise formula may depend on the neutrino mixing angles from the couplings $\bar{\nu}_i\gamma^\mu\nu_j X_\mu$.

The kinetic decoupling of DM from the relativistic particles imprints on the matter power 
spectrum, for which there are two relevant scales~\cite{Loeb:2005pm,Bertschinger:2006nq}: 
the comoving horizon $\tau_{\textrm{kd}}\propto 1/T^{\textrm{kd}}_{\chi}$ and free-streaming length $\left(T^{\textrm{kd}}_{\chi}/m_\chi\right)^{1/2}\tau_{\textrm{kd}}$. For our interested regime, $\tau_{\textrm{kd}}$ is much larger and relevant. Thus $T^{\textrm{kd}}_{\chi}$ can be translated into a cutoff in the power spectrum of  matter density perturbation with 
\[
M_{\textrm{cut}} = \frac{4\pi}{3}\rho_{\textrm{M}}\left(c\tau_{\textrm{kd}}\right)^3\sim 2\times 10^8 \left( \frac{ T^{\textrm{kd}}_\chi }{ \textrm{keV} } \right)^{ -3 } M_{\odot},
\]
where $\rho_{\textrm{M}}$ is the sum of matter densities today, $\rho_{\textrm{CDM}}+\rho_{\textrm{baryon}}$.
Then $M_\textrm{cut}\sim \mathcal{O}(10^9)M_{\odot}$ can be easily obtained for explanation of {\textit{missing satellites problem}} for $\mathcal{O}$(TeV) $\chi$ and $\mathcal{O}$(MeV) $X_\mu$.

\begin{figure}
\includegraphics[width=0.45\textwidth]{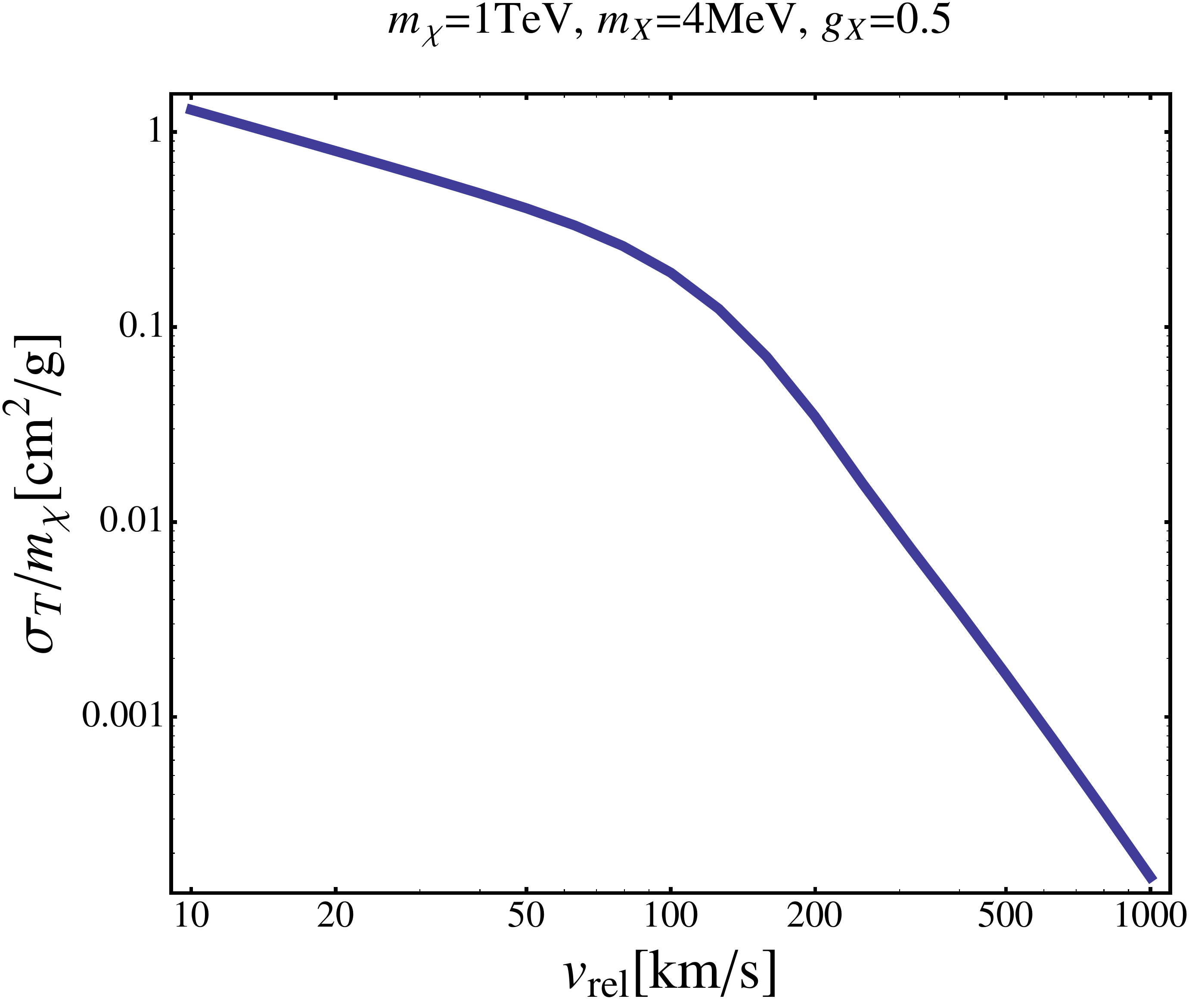}
\caption{$\sigma_T/m_\chi$ as function of relative velocity for $m_\chi=1\mathrm{TeV},m_X=4\mathrm{MeV}$ and $g_X=0.5$.}
\label{fig:sigmaT}
\end{figure}

Because of the light mediator $X_\mu$, the DM self-scattering $\chi \bar{\chi}\rightarrow \chi \bar{\chi}$  can have a large cross section, $\sigma\sim 1$cm$^2/$ at small scales, while relative small values at Milky Way and larger scales. This can flatten the dark halo, decrease the total mass of halo centre and resolve both {\textit {cusp vs. core}} and {\textit{too-big-to-fail}} controversies. 
The quantity that is usually used to describe the efficiency for the DM-DM self-scattering 
is the transfer cross section
\begin{equation*}
\sigma_{T}\equiv \int d\Omega (1-\cos\theta) \frac{d\sigma}{d\Omega} \ . 
\end{equation*}
$\sigma_{T}$ can be easily calculated from Fig.~\ref{fig:DMnu}(a) in the perturbative regime $\alpha_X m_\chi<m_X$ as, 
\begin{eqnarray*} 
&&\sigma_{T}  =  \frac{8\pi}{m_X^{2}}\beta^{2}\left[\ln\left(1+R^{2}\right)-\frac{R^{2}}{1+R^{2}}\right] \ ,
\\
&&\alpha_X=\dfrac{g_{X}^{2}}{4\pi}, \beta  =  \frac{2\alpha_{X}m_X}{m_\chi v_{\textrm{rel}}^{2}} \ ,
R=\dfrac{m_\chi v_{\textrm{rel}}}{m_X},
\end{eqnarray*}
where $v_{\textrm{rel}}$ is the relative velocity of $\chi$ and $\bar{\chi}$. $v_{\textrm{rel}}$ is around $20,200,1000$ km/s for Dwarf galaxies, Milky Way and the galaxy clusters, respectively. In the non-perturbative regime $\alpha_X m_\chi>m_X$, we have~\cite{Feng:2009hw}
\begin{eqnarray*}
\sigma_T = 
\left\{\begin{array}{lc}
\frac{4 \pi}{m_X^2} \beta^2 \ln\left(1+\beta^{-1}\right) & \beta \lesssim 0.2 \\
\frac{8 \pi}{m_X^2} \beta^2 / \left(1+1.5 \beta^{1.65}\right) & \; 0.2 \lesssim \beta \lesssim 1300 \\
\frac{\pi}{m_X^2} \left(\ln \beta+1-\frac{1}{2} \ln^{-1}\beta \right)^2 & \beta \gtrsim 1300
\end{array} \right. 
\end{eqnarray*}
As an illustration, in Fig.~\ref{fig:sigmaT}, we show the case with  $m_\chi=1\mathrm{TeV},m_X=4\mathrm{MeV}$ and $g_X=0.5$, in which
$\sigma_T/m_\chi$ can be achieved properly for Dwarf galaxies with $v_\textrm{vel}\simeq 20$km/s.

\section{Effective Number of Extra Neutrinos}

\begin{figure}[htb]
\includegraphics[width=0.45\textwidth]{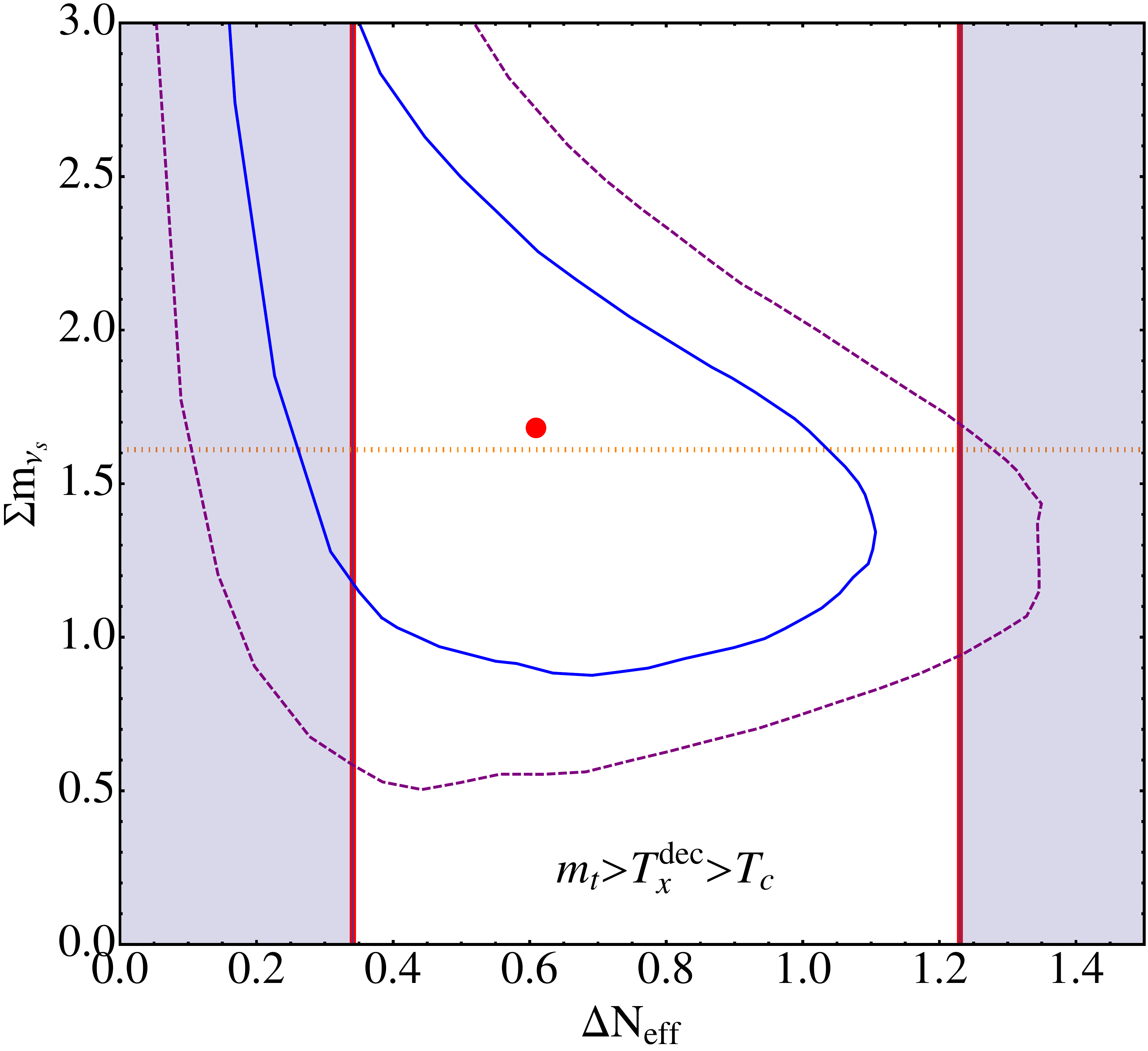}
\caption{The allowed range for $\Delta N_\textrm{eff}$ and $\sum m_{\nu_s}$. The blue(solid) and purple(dashed) contours~\cite{Hamann:2013iba} correspend to the $1\sigma$ and $2\sigma$ for the cosmological data with the best fit point $\Delta N_\textrm{eff}=0.61\pm 0.30,\; m^{\textrm{eff}}_{\textrm{hdm}}=(0.47\pm 0.13)\textrm{ eV}$. The region between two red vertical lines can be achieved in our model. And the horizontal dotted line marks the centre value for $\sum m_{\nu_s}$ from the global fit for neutrino oscillation data in $3+2$ scenario~\cite{globalfitnu}. We use $m_t\simeq 173$GeV and $T_c$ is the confinement-deconfinement transition between quarks and hadrons. See the text for detail.\label{fig:fit}}
\end{figure}

After the decoupling of dark sector from the visible thermal bath, relativistic particles can still contribute to the radiation density.  For 4 light sterile neutrinos, their contributions to 
$\Delta N_{\textrm{eff}}$ can be parametrized as 
\begin{eqnarray}\label{eq:dneff}
\Delta N_{\textrm{eff}}\left(T\right) 
 & = & 4\times \frac{T_{\nu_{s}}^{4}}{T_{\nu_{a}}^{4}} 
 = 4\times\left[\frac{g_{\ast s}\left(T\right)}{g_{\ast s}^{x}\left(T\right)} \times 
   \frac{g_{\ast s}^{x}\left(T\right)T_{\nu_{s}}^{3}}{g_{\ast s}\left(T\right)T_{\nu_{a}}^{3}}\right]^{\frac{4}{3}}\nonumber \\
  & = & 4\times\left[\frac{g_{\ast s}\left(T\right)}{g_{\ast s}^{x}\left(T\right)}\times \frac{g_{\ast s}^{x}\left(T_{x}^{\textrm{dec}}\right)} {g_{\ast s}\left(T_{x}^{\textrm{dec}}\right)}\right]^{\frac{4}{3}},
\end{eqnarray}
where $T$ is the photon temperature, 
and $g_{\ast s}$ counts the total number of relativistic degrees of freedom for entropy 
($g_{\ast s}^{x}$ for dark sector). 
Conservation of entropy density has been used in the last step of the above equations.

When only sterile neutrinos are relativistic at the time just before BBN epoch, we have 
\begin{align*}
& g^{x}_{\ast s} \left(T^{\textrm{dec}}_{x}\right) = 3 + 1 + \frac{7}{8}\times\left(4\times 2\right)=11, \\
& g^{x}_{\ast s} \left(T_{\textrm{bbn}}\right)= \frac{7}{8}\times\left(4\times 2\right)=7.
\end{align*}
The parameter $g_{\ast s} \left(T^{\textrm{dec}}_{x}\right)$ is well-known in SM~\cite{Beringer:1900zz} and depends on the decoupling temperature. For example, $g_{\ast s} \left(T^{\textrm{dec}}_{x}\right) \simeq 72$ for $m_c < T^{\textrm{dec}}_{x} < m_{\tau}$. 
Together with 
\[
g_{\ast s} \left(T_{\textrm{bbn}}\right)= 2 + \frac{7}{8}\times\left(3\times 2 + 2\times 2\right)=\frac{43}{4},
\]
we can get
\begin{equation}
\Delta N_{\textrm{eff}} = 4\times \left[\frac{\frac{43}{4}\times 11}{7\times 72}\right]^{\frac{4}{3}} \simeq 0.579.
\end{equation}
Increasing(decreasing) $T^{\textrm{dec}}_{x}$ gives smaller(larger) $\Delta N_{\textrm{eff}}$ 
due to the changes in $g_{\ast s} \left(T^{\textrm{dec}}_{x}\right)$. 
For instance, if $T^{\textrm{dec}}_{x} > m_t$, we would have $g_{\ast s} \left(T^{\textrm{dec}}_{x}\right) \simeq 107 $ and $\Delta N_{\textrm{eff}}=0.341$. 
If $T_c<T^{\textrm{dec}}_{x} < m_s$, we would have $g_{\ast s} \left(T^{\textrm{dec}}_{x}\right)
 \simeq 41 $ and $\Delta N_{\textrm{eff}}=1.23$. Here $T_c$ is the temperature for 
 confinement-deconfinement transition between quarks and hadrons in QCD.
 
Decoupling temperature lower than $T_c$ would give too large $\Delta N_{\textrm{eff}}\gtrsim 3.96$ and therefore is excluded at high confidence level. The available range for $\Delta N_{\textrm{eff}}$ is the region between two red vertical lines in Fig.~\ref{fig:fit}.


If $X_\mu$ and $H_2$ are also relativistic around BBN time, we have $g_{\ast s}^{x}\left(T\right)=g^{x}_{\ast s}\left(T_{x}^{\textrm{dec}}\right)$ in Eq.~(\ref{eq:dneff}) and additional contributions from the bosonic part
\begin{equation*}
\Delta N^{b}_{\textrm{eff}}=2\times \frac{8}{7}\times \frac{T_{\nu_{s}}^{4}}{T_{\nu_{a}}^{4}}, 
\end{equation*} 
where the factor $2$ accounts for bosonic degrees of freedom normalized to fermonic one, $ {g_b}/{g_{\nu}}$. The ratio of $\Delta N_{\textrm{eff}}$ for two cases is about 
\begin{equation}
{\rm ratio} =\frac{4 \times \left(\frac{11}{7}\right)^\frac{4}{3}}{4 + 2\times \frac{8}{7}}\simeq 1.16.
\end{equation}
So the difference is small and we shall not distinguish two cases in the later discussion.

These extra sterile neutrinos can also be relativistic even at CMB time with 
$T_\gamma \simeq O(1)$ eV and play the role of HDM. Their effects on cosmology can be parametrized by the effective mass  defined as  
\begin{equation}\label{eq:masseff}
m^{\textrm{eff}}_{\textrm{hdm}}\equiv \left(\frac{T_{\nu_s}}{T_{\nu_a}}\right)^3\sum_{\nu_s} m_{v_s}
=\left(\frac{\Delta N_{\textrm{eff}}}{4}\right)^\frac{3}{4}\sum_{\nu_s} m_{v_s},
\end{equation}
where only relativistic sterile neutrinos are summed over.

Sterile neutrino masses can be chosen to fit the neutrino oscillation data. 
We take the face values from the global fit~\cite{Kopp:2011qd, Archidiacono:2012ri, globalfitnu}: for instance, with $3+2$ scenario \cite{globalfitnu} gives $\Delta m^2_{41}=0.46\textrm{~eV}^2$ and $\Delta m^2_{51}=0.87\textrm{~eV}^2$. 
Since $\nu_1$ is massless in our model, we have $m_4\simeq 0.68$~eV and 
$m_5\simeq 0.93$~eV.  Then using Eq.~(\ref{eq:masseff}), we depict the central value 
of $\sum m_{\nu_s}$ as the horizontal dotted line in Fig.~\ref{fig:fit} . 
We can see that cosmological data can be reconciled with neutrino oscillation 
experiments within $1\sigma$ in our model, which is quite remarkable.

The crucial difference between our model and Ref.~\cite{Bringmann:2013vra} is due to 
Eqs.~(\ref{eq:dneff}) and (\ref{eq:masseff}),  because of ``4'' sterile neutrinos in our model. 
Usually, only one sterile neutrino is responsible for $\Delta N_{\textrm{eff}}$ and 
the relation among $m_4$, for which one would have 
\[
m_{\textrm{hdm}}^{\textrm{eff}}=\left(\Delta N_{\textrm{eff}}\right)^{3/4} m_4 . 
\]
Then this is consistent with neutrino oscillation data only at $2\sigma$ level as shown in 
Ref.~\cite{Bringmann:2013vra}.

In the above discussion we have assumed that $\Delta N_{\textrm{eff}}
\left(\textrm{BBN}\right)=\Delta N_{\textrm{eff}}\left(\textrm{CMB}\right)$ for illustration. 
This assumption may not be necessarily true when either oscillation brings all neutrinos 
into equilibrium or some sterile neutrinos are heavy enough such that they become 
non-relativistic at the time before CMB and heat other neutrinos.  In both cases we have  
$\Delta N_{\textrm{eff}}\left(\textrm{CMB}\right)<
\Delta N_{\textrm{eff}}\left(\textrm{BBN}\right)$, and our model predictions are still consistent 
with neutrino oscillation data within $1\sigma$ level.

\section{Further Tests of the Model}

There are a few different ways to test our model. Direct detection of CDM $\chi$ will be 
possible for no vanishing $\sin\epsilon$.  Also, $\chi\bar{\chi}$ will annilhilate to two 
$X_\mu$s,   which in turn decay into sterile neutrinos immediately. 
These high energy sterile neutrinos can oscillate to active neutrinos which can be 
detected by neutrino telescropes, such as IceCube, whose current limit on 
$\langle \sigma v\rangle$ is around $10^{-22}$cm$^3/$s~\cite{Aartsen:2013mla}. 
Taking into account boost factors due to light mediators in our model,  future detection 
of these neutrino flux will be possible.  Since we have more sterile neutrino species 
than other models, 
oscillation experiments could also be used to test the model even though 
this depends on the exact mixing angles and mass spectrum.  

\section{summary}
In this paper, we have proposed a {\textit{ultraviolet}} complete renormalizable model 
for self-interacting CDM and sterile neutrinos that can accommodate the cosmological 
data and neutrino oscillation experiments simultaneously within $1\sigma$ level.  
The model is based on a dark sector with  local $U(1)_X$ dark gauge symmetry 
that is spontaneously broken at $\mathcal{O}$(MeV) scale. The resulting 
$\mathcal{O}$(MeV) gauge boson (dark photon) can mediate a DM 
self-scattering cross section around $\sigma \sim 1$cm$^2/$g which is of right order to 
resolve two issues for CDM  at small cosmological scales, {\textit {cusp vs. core}} and 
{\textit{too-big-to-fail}}. 

In our model,  two light RH gauge singlet neutrinos ($N_{i=1,2}$)  can mix with 
a dark fermion $\psi$ and therefore can interact with DM through the new dark gauge 
boson. The relics of these sterile neutrinos serve as the hot dark matter with a right 
amount of $\Delta N_{\textrm{eff}}$ (see Fig.~1), which relieves the tension between Planck and BICEP2. 
The masses of these sterile neutrinos are consistent with neutrino oscillation experiments 
within  $1\sigma$.  Meanwhile, the interaction between DM and sterile neutrino delays 
the DM's kinetic decouple to sub-keV temperature and induces a lower cut-off in the 
primordial matter power spectrum, resolving the {\textit{missing satellites problem}}.
The model could be tested further  by (in)direct detection of CDM $\chi$, and also through
neutrino oscillation experiments if favorable parameters are realized in Nature.

\begin{acknowledgments} 
We are grateful to Wan-Il Park for useful discussions. 
This work is supported in part by National Research Foundation of Korea (NRF) Research 
Grant 2012R1A2A1A01006053 , and by the NRF grant funded by the Korea 
government (MSIP) (No. 2009-0083526) through  Korea Neutrino Research Center 
at Seoul National University (PK).
\end{acknowledgments}

\end{document}